\documentclass[conference]{IEEEtran}
\IEEEoverridecommandlockouts
\usepackage{cite}
\usepackage{caption}
\usepackage{subcaption} 
\usepackage{diagbox}
\usepackage{float}
\usepackage{makecell} 
\usepackage{array}
\usepackage{svg}
\usepackage{tabularx}
\usepackage{epstopdf}
\usepackage{amsmath, amsfonts, amssymb}
\usepackage{algorithm}
\usepackage{algorithmic}
\usepackage{fullpage} 
\usepackage{graphicx}
\usepackage{subcaption}
\usepackage{booktabs}
\usepackage{afterpage}
\usepackage{textcomp}
\usepackage{xcolor}
\def\BibTeX{{\rm B\kern-.05em{\sc i\kern-.025em b}\kern-.08em
    T\kern-.1667em\lower.7ex\hbox{E}\kern-.125emX}}
\usepackage{svg}
\usepackage{indentfirst}
\usepackage{amsmath}
\usepackage{amssymb}
\usepackage{microtype}
\usepackage{geometry}
\usepackage{amsthm}
\usepackage{mathtools}
\usepackage{stmaryrd}
\usepackage{textcomp}
\usepackage{tikz}
\usetikzlibrary{arrows,automata,positioning,shapes,patterns,arrows.meta,calc}
\usepackage{pifont}
\usepackage{float}
\usepackage[capitalise]{cleveref}
\usepackage{xspace}
\usepackage{graphicx}
\usepackage{listings, multicol, multirow}
\usepackage{caption}
\usepackage{wrapfig}
\usepackage{array}
\usepackage{makecell}
\usepackage{booktabs, tabularx}
	\newcolumntype{Y}{>{\centering\arraybackslash}X}

\makeatletter
\newcommand\fs@betterruled{%
  \def\@fs@cfont{\bfseries}\let\@fs@capt\floatc@ruled
  \def\@fs@pre{\vspace*{5pt}\hrule height.8pt depth0pt \kern2pt}%
  \def\@fs@post{\kern2pt\hrule\relax}%
  \def\@fs@mid{\kern2pt\hrule\kern2pt}%
  \let\@fs@iftopcapt\iftrue}
\floatstyle{betterruled}
\restylefloat{algorithm}
\makeatother

\include{macros}

\theoremstyle{definition}

\pgfdeclarelayer{background}
\pgfsetlayers{background,main}

\begin{document}
%
\title{DPFedBank: Crafting a Privacy-Preserving Federated Learning Framework for Financial Institutions with Policy Pillars}

\author{
\IEEEauthorblockN{$^{1}$Peilin He$^{\dagger}$, $^{2}$Chenkai Lin$^{\dagger}$, $^{3}$Isabella Montoya$^{\dagger}$}\\
\IEEEauthorblockA{
\textit{$^{1}$Department of Informatics and Networked Systems, University of Pittsburgh, USA}\\
\textit{$^{2}$School of Public Policy and Management, Carnegie Mellon University, USA}\\
\textit{$^{3}$Information Network Institute, Carnegie Mellon University, USA}\\
Emails: peilin.he@pitt.edu$^{1}$, chenkai2@andrew.cmu.edu$^{2}$, imontoya@andrew.cmu.edu$^{3}$
}
}


%


\maketitle

\begin{abstract}
In recent years, the financial sector has faced growing pressure to adopt advanced machine learning models to derive valuable insights while preserving data privacy. However, the highly sensitive nature of financial data presents significant challenges to sharing and collaboration. This paper presents DPFedBank, an innovative framework enabling financial institutions to collaboratively develop machine learning models while ensuring robust data privacy through Local Differential Privacy (LDP) mechanisms. DPFedBank is designed to address the unique privacy and security challenges associated with financial data, allowing institutions to share insights without exposing sensitive information. By leveraging LDP, the framework ensures that data remains confidential even during collaborative processes, providing a crucial solution for privacy-aware machine learning in finance. We conducted an in-depth evaluation of the potential vulnerabilities within this framework and developed a comprehensive set of policies aimed at mitigating these risks. The proposed policies effectively address threats posed by malicious clients, compromised servers, inherent weaknesses in existing Differential Privacy-Federated Learning (DP-FL) frameworks, and sophisticated external adversaries. Unlike existing DP-FL approaches, DPFedBank introduces a novel combination of adaptive LDP mechanisms and advanced cryptographic techniques specifically tailored for financial data, which significantly enhances privacy while maintaining model utility. Key security enhancements include the implementation of advanced authentication protocols, encryption techniques for secure data exchange, and continuous monitoring systems to detect and respond to malicious activities in real-time. Furthermore, DPFedBank features an adaptive threat detection module that leverages anomaly detection to identify potential breaches more effectively. Our findings demonstrate that the DPFedBank framework not only significantly improves the privacy, security, and resilience of federated learning systems in the financial sector but also provides an enhanced balance between data utility and privacy, thereby fostering safer and more effective collaboration among financial entities.
\end{abstract}


%
\IEEEpeerreviewmaketitle

\section{Introduction}
\label{1}
Modern financial institutions are under increasing pressure to leverage advanced machine learning techniques for valuable insights while safeguarding sensitive data. Stringent data protection requirements in the financial sector create significant barriers to data sharing and collaboration. To address these challenges, institutions must adopt robust privacy-preserving strategies that mitigate the unique risks associated with financial data. The proposed framework, \textit{DPFedBank}, is a \textit{Federated Learning (FL)}-based system that enables financial institutions to collaboratively develop machine learning models while upholding stringent data privacy standards. It employs advanced techniques such as Local Differential Privacy (LDP) and rigorous policy implementation to protect against potential threats.

\subsection{Federated Learning as a Promising Solution}
Among various privacy-preserving techniques, \textit{Federated Learning (FL)} has emerged as a potential solution for secure collaborative modeling in finance. FL allows multiple institutions to train a shared model without requiring the transfer of raw data to a central server. By enabling data holders to contribute to model development while keeping their data local, FL reduces privacy risks, making it suitable for sectors like finance where data confidentiality is critical \cite{mammen2021federated}. The DPFedBank framework extends this capability by incorporating LDP mechanisms tailored for financial data, ensuring that privacy is maintained even during collaborative processes.

FL aims to achieve core privacy objectives such as decentralized data storage, secure data communication, and data confidentiality. In a federated learning system, each financial institution acts as an independent worker, analogous to a distributed computing environment with enhanced privacy protections. Through encrypted communication between the central server and participating institutions, machine learning can be conducted without exposing sensitive records \cite{li2020review}. However, despite its advantages, FL still encounters specific challenges in practical applications.

\subsection{Challenges in Federated Learning}
A major challenge in applying FL to financial institutions is balancing \textit{privacy, performance, and regulatory compliance}. While encryption-based methods provide robust privacy, they can introduce computational inefficiencies, increasing latency. To mitigate this, techniques like \textit{differential privacy (DP)} have been employed, where noise is introduced at the client level before model updates. The \textit{NbAFL method} proposed by Wei et al. \cite{wei2020federated} demonstrates the benefits of client-side noise addition for protecting privacy while maintaining model accuracy. This approach is particularly beneficial for financial institutions with diverse data distributions.

However, DP presents fairness concerns as privacy guarantees may vary for clients with unique or rare data characteristics. Clients with such data may receive weaker privacy protection, raising ethical and regulatory issues regarding equal privacy guarantees across the system.

\subsection{Vulnerabilities in Federated Learning}
The federated learning process in financial institutions exposes several vulnerabilities across different stages, including client selection, global model distribution, local training, aggregation, and model updates \cite{bouacida2021vulnerabilities}. These vulnerabilities can compromise privacy and integrity if not addressed adequately:

1) \textit{Malicious Clients}: Financial institutions could encounter clients that intentionally modify model parameters or introduce misleading updates, thereby degrading the global model's quality or extracting sensitive information from other participants.

2) \textit{Aggregator Threats}: If an adversary gains control of the aggregation role, they could manipulate or selectively ignore updates from legitimate institutions, resulting in a biased or corrupted global model.

3) \textit{Compromised Server}: A compromised central server could interfere with model updates or perform unauthorized data analysis, potentially violating data protection regulations.

4) \textit{External Adversaries}: External threats, such as cyberattacks on communication channels, could expose sensitive financial data during transmission or alter the model updates.

\subsection{Holistic Approach to Mitigating Risks}
To address these vulnerabilities, the DPFedBank framework adopts a \textit{holistic and policy-driven approach} that includes multiple security layers and privacy-enhancing techniques. The policy pillars within DPFedBank focus on protecting data confidentiality, model integrity, and regulatory compliance across all stages of federated learning \cite{kairouz2021advances}. These pillars encompass strategies for securing client interactions, server operations, output models, and stakeholder responsibilities, ensuring that privacy is preserved throughout the federated learning process.

\subsection{Contributions of This Paper}
This paper introduces \textit{DPFedBank}, a policy-driven federated learning framework that enhances privacy protection in financial institutions through tailored privacy-preserving techniques. By addressing critical vulnerabilities and integrating adaptive LDP mechanisms, DPFedBank ensures a balance between data privacy, model utility, and regulatory requirements. The proposed framework includes advanced security measures such as cryptographic protocols, secure data exchange, and continuous monitoring to detect and respond to threats in real-time. Our evaluations demonstrate that DPFedBank significantly improves the resilience of federated learning in the financial sector, fostering more secure and effective collaboration among financial institutions.

\section{Background}
\label{2}

\subsection{Privacy-preserving of Federated Learning}
\label{2.1}
Federated learning is a decentralized machine learning technique that trains a model across multiple devices by sending data to a central server. The architecture of this system has every device processes its local data to then update based on that date. Afterwards, this is aggregated by a central server to enhance the performance of the globally shared model. This differs from traditional centralized machine learning which stores and collects data in one location. With this difference federated learning inherently maintains user privacy, by storing data locally, while allowing collaborative model development \cite{FederatedLearning}. Differential privacy  can be used to further preserve user privacy by introducing statistical noise into an individual model before transmission to the server. This then enhances an individual's privacy because changes cannot be linked to any particular individual’s data, allowing for accurate global model training while still offering a mathematical guarantee of privacy. \cite{Han2023FederatedLD} Additionally, the field has be been improve on this model to protect individual privacy. A recent implementation of differential privacy provides a  privacy budget in order to regulate the amount of information that can be disclosed about any individual within a datase\cite{Shen2023ADP}t. This regulation is place to enforce strict data access to protect individual privacy. 

By combining federated learning and differential privacy a framework for preserving user privacy in  machine learning starts to form, which can be crucial in domains like healthcare and finance.This allows organizations to leverage machine learning without jeopardizing individual privacy by limiting the centralization of raw data and implementing these privacy-preserving methods for model changes. Additionally, this strategy aligns with current data privacy laws like GDPR and HIPAA, which place a strong emphasis on limiting data exposure and giving individuals control over their personal data.

\subsection{Challenges of in Differential Privacy and Federated Learning}
\label{2.2}
While implementing differential privacy in federated learning represents a significant advancement, it still presents challenges that should be addressed. In federated learning the effectiveness of communication is a challenge; updating the model often from several devices can be resource-intensive.This can be strenuous for mobile devices with constrained bandwidth and battery life\cite{Drainakis2020FederatedVC}. Additionally, heterogeneous data can impact the model performance, the non-IID (Independent and Identically Distributed) distribution of data across devices might make model training more challenging. Since the variability of the data is necessary to ensure a mathematical guarantee of privacy the aggregated model must function well even with variations in the local data distributions\cite{Nilsson2018APE}. Another challenge for these systems is model security, federated learning systems can be targeted for attacks like inference attacks and model poisoning. In an inference attack is when attackers analyze the machine learning model's outputs or parameters, to retrieve private information about specific individuals. Model Poisoning is an attack where attackers purposefully provide misleading data into the training process of a machine learning model. This allows attackers to degrade the integrity of the model by changing the training data or model updates.  Which can be used to control the model's behavior, reduce performance, or provide biased outputs\cite{DBLP:journals/corr/abs-1911-11815}. These attacks have the ability to compromise user privacy by disclosing information about the training data even within federated learning systems that implement differential privacy.

\subsection{Policy Perspective of Data Privacy}
\label{2.3}

Protecting personal information from unauthorized access, misuse, or disclosure is referred to as data privacy. This concept is crucial, as any weaknesses in these protections can lead to harm to individuals\cite{PrivacyHarm}. Privacy policies help dictate how organizations uphold data privacy based on rules to gather, store, and use personal data. These policies aid in setting standards for protecting user’s rights when organizations are handling user data. 

A prominent example of one of the data privacy enhancing policies is through regulations that are set in the European Union's General Data Protection Regulation (GDPR). This piece of policy sets rules for user data protection for organizations . The GDPR mandates that companies must receive users' explicit consent before processing their personal data and then explain to them how their data will be used \cite{GDPR}. This law provides privacy safeguards and makes organizations responsible for their data practices. Solove has shown that these legal frameworks are necessary to manage the intricacies of data privacy\cite{PrivacyLaw}. In the changing digital environment, we need strong regulations that put user rights and data security first even with evolving technologies. 

To complement privacy laws like the GDPR, technical frameworks like IBM’s Federated Learning aim to provide privacy-enhancing strategies to safeguard sensitive data in systems that utilize federated learning. IBM’s Federated Learning employs secure multiparty computation (SMC) and differential privacy to improve privacy in federated learning systems through a hybrid approach. SMC employs encrypted communication between users, this prevents raw data from being exposed during training.  Differential privacy guarantees that individual data points cannot be identified by introducing noise into the model updates. By implementing these two strategies they can reduce the risk of data inference attacks while maintaining the functionality of the trained models \cite{10.1145/3338501.3357370}.This hybrid model shows how technical advancements can support privacy laws such as GDPR to offer privacy protections for sensitive information.

\section{Proposed DPFedBank Framework}
\label{3}
The framework of DPFedBank is presented in this section. We first demonstrate the overview of DPFedBank. Then we detail three key modules: (1) data distribution for local clients (2) local differential privacy implementation, and (3) the federated learning architecture integration.
\subsection{Threat Modeling}
\label{3.1}
In the rapidly evolving landscape of financial technology, the DPFedBank framework emerges as a pivotal solution for enabling secure and privacy-preserving collaborative machine learning among financial institutions. At the heart of the DPFedBank framework lie several critical assets, including sensitive financial data, global model parameters \(\theta_g\), and locally perturbed model updates \(\tilde{\theta}_n\) transmitted from clients to the aggregator. These assets are highly attractive targets for various threat actors, including malicious clients, honest-but-curious servers, external adversaries, and insider threats. Malicious clients may attempt to manipulate the federated learning process by submitting poisoned or biased model updates, thereby degrading the performance or skewing the outcomes of the global model. Honest-but-curious servers, while ostensibly following protocol, may seek to infer sensitive information from the aggregated data or individual model updates. External adversaries might exploit vulnerabilities in communication channels to intercept or tamper with data transmissions, while insider threats pose risks from within the organizations involved, where authorized personnel might misuse their access privileges to extract or compromise sensitive information. Understanding the motivations and capabilities of these threat actors is crucial in designing effective countermeasures \cite{shokri2015privacy}.

One of the foremost threats to the DPFedBank framework is data inference attacks, wherein adversaries aim to reconstruct sensitive client data from the global model or the aggregated updates. Techniques such as model inversion and membership inference can potentially extract proprietary financial information, undermining the privacy guarantees the framework seeks to uphold \cite{fredrikson2015model}. To mitigate this risk, the implementation of Local Differential Privacy (Local-DP) is instrumental. By introducing carefully calibrated noise into the local model updates before transmission, Local-DP ensures that the contribution of any single data point is obfuscated, thereby preventing precise inference attacks \cite{dwork2014algorithmic}. Additionally, employing robust cryptographic protocols, such as secure multi-party computation and homomorphic encryption, can further enhance data privacy by enabling computations on encrypted data without exposing the underlying information \cite{abadi2016deep}.

Another significant threat vector is model poisoning attacks, where malicious clients inject misleading or adversarial updates into the federated learning process with the intent of degrading model performance or embedding backdoors \cite{bagdasaryan2020backdoor}. These attacks can be particularly insidious as they exploit the inherent trust in the federated learning paradigm, where the central aggregator relies on the integrity of client-submitted updates. To counteract model poisoning, the DPFedBank framework incorporates server-side verification mechanisms, including cryptographic proofs and anomaly detection algorithms, to authenticate and validate the integrity of incoming model updates \cite{geyer2017differentially}. Techniques such as robust aggregation methods (e.g., median or trimmed mean) can also be employed to minimize the influence of outlier updates, thereby ensuring that the aggregated global model remains resilient against malicious perturbations \cite{lyu2022privacy}. Furthermore, implementing client reputation systems can help in identifying and isolating clients that consistently submit suspicious or suboptimal updates, thereby maintaining the overall health of the federated learning ecosystem.

Insider threats, wherein authorized personnel within client institutions or the central server exploit their access privileges, pose a unique challenge to the DPFedBank framework. These threats are particularly pernicious as insiders often possess legitimate access to sensitive components of the system, making malicious activities harder to detect and prevent. To mitigate insider threats, the framework enforces strict access control policies \cite{sandhu1993lattice,samarati2000access}, ensuring that individuals have the minimum necessary privileges to perform their roles.

In conclusion, the threat modeling of the DPFedBank framework underscores the multifaceted security challenges inherent in deploying federated learning within the highly regulated and sensitive domain of financial institutions. By meticulously identifying potential threats such as data inference attacks, model poisoning, MitM attacks, and insider threats, and by implementing a comprehensive suite of mitigation strategies—including Local-DP, secure communication protocols, server-side verification, and stringent access controls—the framework can achieve a robust security posture. Continuous evaluation and adaptation of these measures are essential to address the evolving threat landscape and to maintain the confidentiality, integrity, and availability of both the data and the machine learning models. Ultimately, the DPFedBank framework exemplifies a balanced approach to harnessing the benefits of federated learning while steadfastly safeguarding against its attendant security risks.

\subsection{Overall Structure}
\label{3.2}
\begin{figure*}[t]
    \centering
    \includegraphics[width=\linewidth]{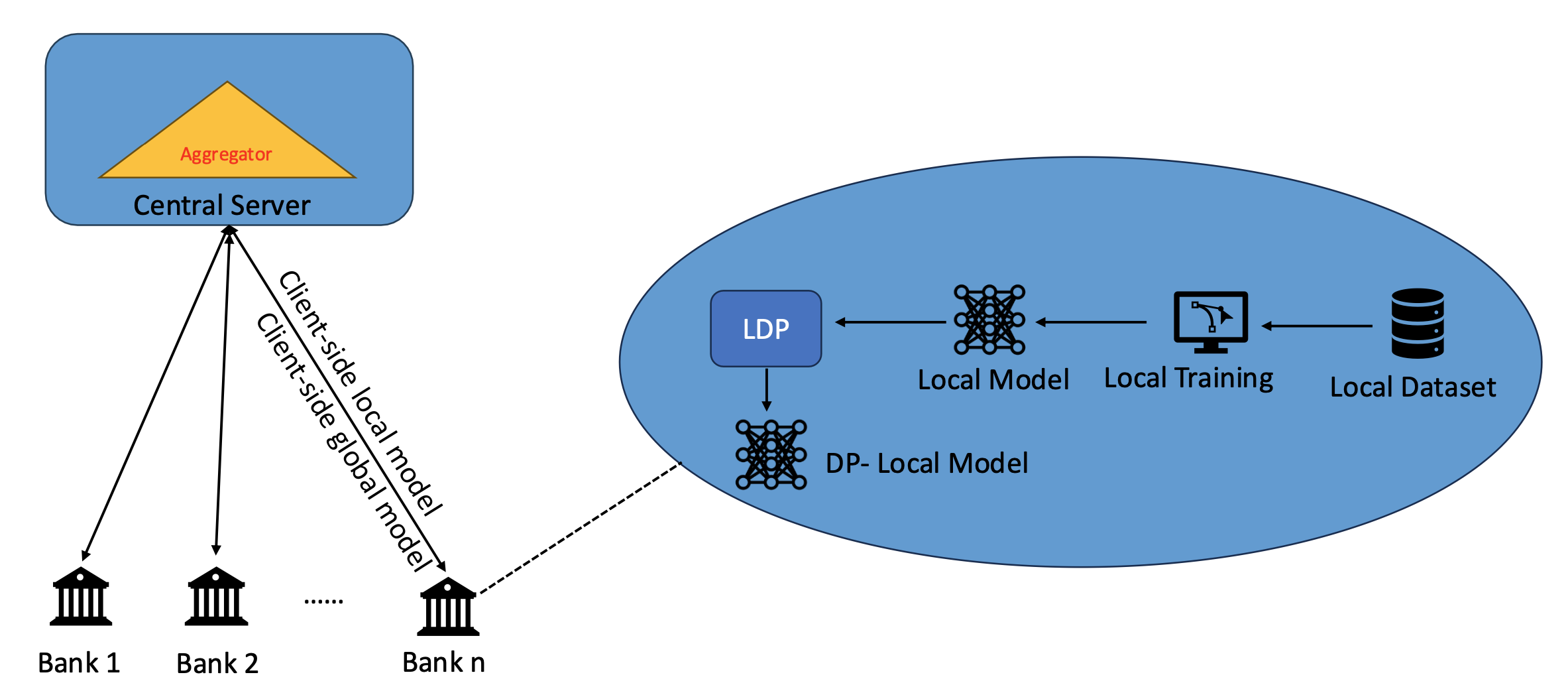}
    \caption{DPFedBank Overview}
    \label{fig:overview}
\end{figure*}

The proposed DPFedBank framework enables multiple financial institutions to collaboratively train a machine learning model without sharing raw data \cite{mcmahan2017communication}, ensuring strong privacy guarantees through Local Differential Privacy (Local-DP) \cite{duchi2013local, triastcyn2019federated, dwork2006differential, dwork2014algorithmic}. Each institution perturbs its model updates locally before sending them to a central aggregator, which computes the global model. This design addresses the unique privacy and regulatory challenges in the financial sector. Refer to \cref{fig:overview} 
for a representation of the DPFedBank architecture. Our DPFedBank framework considers each client as a financial institution, (eg. bank, credit union, etc.), these types of institutions usually have sensitive data, and store their clients' personal information, and that information is typically present as non-IID data \cite{zhao2018federated,zhu2021federated,ma2022state}, which poses a privacy risk as it can create imbalances in local datasets, making it easier to infer individual users’ information. This imbalance leads to the potential leakage of sensitive attributes, since patterns unique to certain groups may be overrepresented. As a result, adversaries can exploit these patterns to identify or extract private details such as the data poisoning attack \cite{singh2023fair} or the label-flipping attack \cite{shen2023privacy} from specific clients, undermining the effectiveness of privacy-preserving mechanisms like differential privacy.

\textbf{DPFedBank Workflow}. The DPFedBank framework operates through an iterative process consisting of several well-defined steps. Each iteration aims to refine the global model while preserving the privacy of individual clients' data. All clients in the network perform forward propagation on their client-side model in parallel, although each client has different levels of sensitive information, they can train their model locally without sharing raw data. Suppose client \(n\) trains its data using dataset \(D_n\), the trained model passes through the LDP mechanism, the LDP is carefully tuned, balancing the trade-offs for model accuracy and privacy budgets. Then the noise version of the local model is transferred to the aggregator, waiting for aggregating all the clients' models, and processes to the central server, the central server distributes the updated model as a one-round global model to all the clients, and all the clients can choose to use the global model or carry on training.

\textbf{Variants of DPFedBank Framework}. There can be developed a variety of DPFedBank architecture, here we only discuss the Vallina DPFedBank framework.

\textbf{System components}. The system includes several components: 
\begin{itemize}
    \item Clients (Financial Institutions), these are entities with local datasets  \(D_n\), where \(n\) denotes the client index. Clients have varying levels of sensitive information, and clients perform local model training.
    \item Local Model Training Module: Allows clients to train models on local data without sharing raw data, training using local forward propagation function \(F\), where  \(\ell(\cdot)\) is the loss function.
    \begin{equation} 
        \min_{\theta_n} F_n(\theta_n) = \frac{1}{|D_n|} \sum_{(x_i, y_i) \in D_n} \ell(\theta_n; x_i, y_i)
    \end{equation}      
    \item Local Differential Privacy Mechanism (LDP Mechanism): Perturbs the trained local models to ensure privacy. Here we choose to add Gaussian noise to the local model parameters:
    \begin{equation}   
    \label{noise}
        \tilde{\theta}_n = \theta_n + \mathcal{N}(0, \sigma^2 I)
    \end{equation}
    where  \(\sigma\)  is the standard deviation calibrated to satisfy  (\(\epsilon-\delta\))-Local DP. An algorithm  \(\mathcal{M}\)  satisfies  \(\epsilon-\delta\)-Local DP if for all possible outputs  \(y\)  and any two inputs  \(x\)  and  \(x{\prime}\):
    \begin{equation}
    \label{dp}
        \Pr[\mathcal{M}(x) = y] \leq e^\epsilon \Pr[\mathcal{M}(x{\prime}) = y] + \delta
    \end{equation}
    If the privacy budgets surpass the threshold \(\delta\), the LDP mechanism will be considered broken. Then the noise calibration is important, we set the noise scale \(\sigma\) to \(\sigma \geq \frac{\Delta}{\epsilon}\), where  \(\Delta\)  is the sensitivity of the local model. The sensitivity  \(\Delta\)  of the local model  \(\theta_n\)  is defined as:
    \begin{equation}
        \Delta = \max_{D_n, D_n{\prime}} \| \theta_n(D_n) - \theta_n(D_n{\prime}) \|_2
    \end{equation}
    where  \(D_n\)  and  \(D_n{\prime}\)  differ by one data point. By the properties of the Gaussian distribution \cite{dwork2009differential}, the probability density function satisfies the required privacy condition.
 
    \item Aggregator: Collects the perturbed models from clients and performs aggregation. The aggregator computes:
    \begin{equation}
    \label{agg}
        \theta_{\text{agg}} = \frac{1}{N} \sum_{n=1}^N \tilde{\theta}_n
    \end{equation}
    where N is the number of clients in the networks.
    \item Central Server: Processes the aggregated model  \(\theta_{\text{agg}}\), updates the global model  \(\theta_g\) and Distributes  \(\theta_g\)  to clients.
\end{itemize}

\textbf{DPFedBank Framework Construction}.
In \cref{alg:client_ldp} and \cref{alg:server_aggregation}, we define all clients (financial institutions) \(n\) in the network to perform forward propagation \cite{li2020federated} on their local models using their datasets \(D_n\) containing sensitive data. Each client computes the local model parameters \(\theta_n\). Clients pass their trained models  \(\theta_n\)  through the LDP mechanism \cite{dwork2006differential,dwork2014algorithmic} to get the perturbed model:  \(\tilde{\theta}_n\) = \(\mathcal{M}(\theta_n)\). The LDP mechanism is carefully tuned to balance model accuracy and privacy budgets. Then, clients transfer the noisy versions of the local models  \(\tilde{\theta}_n\) to the aggregator. The aggregator waits to collect all clients’ perturbed models, aggregates models using  \(\theta_{\text{agg}} = \mathcal{A}(\{ \tilde{\theta}n \}{n=1}^N)\) \cite{bonawitz2017practical}. Then the aggregated model \(\theta_{\text{agg}}\) is processed by the central server, and the server updates the global model  \(\theta_g\)  based on  \(\theta_{\text{agg}}\). The central server distributes the updated global model  \(\theta_g\)  to all clients, clients can choose to use the global model or continue training their local models. Note that, we should be careful that the expected error (difference) between the aggregated model and the true model without noise:
\begin{equation}
   \mathbb{E}[\| \theta_{\text{agg}} - \theta^* \|_2^2] = \frac{\sigma^2}{N}
\end{equation}
where  \(\theta^*\)  is the true aggregated model without noise. 

\begin{algorithm}[t]
\caption{Client-Side Training and Local Differential Privacy Mechanism in DPFedBank}
\label{alg:client_ldp}
\begin{algorithmic}[1]
\REQUIRE
\begin{itemize}
    \item Local dataset $D_n$
    \item Received global model parameters $\theta^{(t)}$ from the central server
    \item Privacy parameter $\epsilon$
    \item Learning rate $\eta$
    \item Number of local epochs $E$
\end{itemize}
\ENSURE
\begin{itemize}
    \item Perturbed local model update $\tilde{\theta}_n^{(t)}$
\end{itemize}
\STATE \textbf{Initialize} local model parameters: $\theta_n^{(t,0)} \leftarrow \theta^{(t)}$
\FOR{each local epoch $e = 1$ to $E$}
    \FOR{each mini-batch $B \subset D_n$}
        \STATE Compute gradient: $g_n^{(e)} \leftarrow \nabla F_n(\theta_n^{(t,e-1)}; B)$
        \STATE Update local model parameters:
        \[
        \theta_n^{(t,e)} \leftarrow \theta_n^{(t,e-1)} - \eta \, g_n^{(e)}
        \]
    \ENDFOR
\ENDFOR
\STATE Compute local model update:
\[
\Delta \theta_n^{(t)} \leftarrow \theta_n^{(t,E)} - \theta^{(t)}
\]
\STATE \textbf{Local Differential Privacy Mechanism:}
\STATE Compute sensitivity:
\[
\Delta \leftarrow \max_{D_n, D_n'} \left\| \Delta \theta_n^{(t)}(D_n) - \Delta \theta_n^{(t)}(D_n') \right\|_2
\]
\STATE Set noise scale:
\[
\sigma \leftarrow \frac{\Delta}{\epsilon}
\]
\STATE Generate noise vector:
\[
b_n^{(t)} \sim \mathcal{N}(0, \sigma^2 I)
\]
\STATE Apply LDP mechanism:
\[
\tilde{\theta}_n^{(t)} \leftarrow \Delta \theta_n^{(t)} + b_n^{(t)}
\]
\STATE Securely send $\tilde{\theta}_n^{(t)}$ to the aggregator
\end{algorithmic}
\end{algorithm}

\begin{algorithm}[t]
\caption{Server-Side Aggregation and Global Model Update in DPFedBank}
\label{alg:server_aggregation}
\begin{algorithmic}[1]
\REQUIRE
\begin{itemize}
    \item Received perturbed local updates $\tilde{\theta}_n^{(t)}$ from all clients $n = 1, 2, \dots, N$
    \item Current global model parameters $\theta^{(t)}$
\end{itemize}
\ENSURE
\begin{itemize}
    \item Updated global model $\theta^{(t+1)}$
\end{itemize}
\STATE \textbf{Aggregation:}
\STATE Compute aggregated update:
\[
\Delta \theta_{\text{agg}}^{(t)} \leftarrow \frac{1}{N} \sum_{n=1}^N \tilde{\theta}_n^{(t)}
\]
\STATE Update global model:
\[
\theta^{(t+1)} \leftarrow \theta^{(t)} + \Delta \theta_{\text{agg}}^{(t)}
\]
\STATE Distribute updated global model $\theta^{(t+1)}$ to all clients
\end{algorithmic}
\end{algorithm}

\subsection{Privacy Protection Analysis}
\label{3.3}
\textbf{Client-Side Privacy Protection}. In the DPFedBank framework, client-side privacy protection is paramount, as each participating financial institution holds sensitive and proprietary data that must remain confidential. The primary mechanism employed to ensure client-side privacy is Local Differential Privacy (Local-DP). Local DP allows clients to add noise to their data or model updates before any information leaves their local environment, ensuring that the central server or any other external entity cannot infer sensitive details from the shared data. As we can refer to \cref{noise}, the noise scale  \(\sigma\) is determined based on the desired privacy parameter \(\epsilon\)  and the sensitivity  \(\Delta\)  of the model updates. By adding this noise, the client’s update becomes differentially private, meaning that any single data point’s influence on the model update is obscured. This mechanism ensures that even if an adversary gains access to the perturbed updates, they cannot accurately infer any individual data point from the client’s dataset. Clients perform a sensitivity analysis to determine the maximum change in the model update that could result from the addition or removal of a single data point.This sensitivity  \(\Delta\)  is crucial for calibrating the noise scale  \(\sigma\):
\begin{equation}
        \sigma = \frac{\Delta}{\epsilon} \sqrt{2 \ln \left( \frac{1.25}{\delta} \right)}
\end{equation}
Here,  \(\epsilon\)  controls the privacy-utility trade-off, and  \(\delta\)  is a small parameter representing the probability of the privacy guarantee being breached. By carefully selecting  \(\epsilon\)  and  \(\delta\) , clients can balance the level of privacy protection with the utility of the model updates.

\textbf{Server-Side Privacy Protection}. On the server side, the DPFedBank framework focuses on safeguarding the aggregated model and preventing any potential leakage of sensitive information during the aggregation and distribution processes. The server plays a critical role in updating the global model while ensuring that individual clients’ contributions remain confidential. The server employs secure aggregation protocols to combine the perturbed model updates  \(\tilde{\theta}_n\)  received from clients. Since each client’s update is already differentially private due to the Local-DP mechanism, the aggregation process further obscures individual contributions. The server computes the aggregated update as referred to \cref{agg}. By averaging the noisy updates, the server reduces the variance introduced by the noise, enhancing the utility of the global model without compromising privacy. The law of large numbers ensures that as the number of clients  \(N\)  increases, the aggregated noise diminishes, improving model accuracy.

\section{Policy and Regulation Development}
\label{5}
The DPFedBank framework presents a robust solution for collaborative machine learning among financial institutions while preserving data privacy through Local Differential Privacy (LDP). However, the deployment of such a system necessitates the formulation of comprehensive policies and regulations to address potential vulnerabilities, such as malicious clients, inherited flaws from Differential Privacy-Federated Learning (DP-FL) frameworks, compromised servers, and threats from external adversaries. Below, we propose novel policies and regulations tailored to the DPFedBank framework, drawing inspiration from existing guidelines but extending them to meet the unique challenges posed by this system.

\subsection{Policy for Mitigating Malicious Clients}
 1) Implement a stringent authentication mechanism using digital certificates and multi-factor authentication to verify the identity of each participating financial institution. This reduces the risk of unauthorized entities joining the network. Similar authentication practices are discussed in \cite{bonawitz2017practical}, emphasizing secure aggregation protocols in federated learning. 2) Inside the system, deploying real-time monitoring tools to analyze the statistical properties of clients' model updates is crucial. Any deviations from expected patterns, such as abnormal parameter distributions or inconsistent training behaviors, can indicate malicious activity. This approach is inspired by techniques in \cite{kairouz2021advances}, which emphasize monitoring and robustness in federated learning systems. 3) For each institution, establishing a reputation system where clients accumulate trust scores based on their compliance with protocols and contribution to model accuracy should be required. Clients exhibiting suspicious behavior may have reduced privileges or be subject to additional scrutiny. This concept extends the ideas presented in \cite{li2020federated} regarding fairness and accountability in federated learning. 4) Also, to create a secure execution environment \cite{sabt2015trusted}, clients should be required to perform local computations within secure enclaves, such as Intel SGX \cite{costan2016intel}, to prevent tampering with code or data. This ensures the integrity of the local training process, aligning with practices outlined in \cite{hunt2018chiron} exploring privacy-preserving machine learning using secure hardware.

 \subsection{Policy for Addressing Heritage Flaws in DP-FL Frameworks}
 1) Conduct regular third-party security audits that include formal verification methods to mathematically prove the correctness and security properties of the LDP mechanisms and aggregation functions. This proactive approach aims to identify and mitigate inherited vulnerabilities from existing DP-FL frameworks, building upon principles from \cite{abadi2016deep} who emphasize rigorous analysis in differential privacy implementations. 2) Design the DPFedBank framework with a modular architecture that allows individual components to be updated independently. This facilitates quick responses to discovered flaws without disrupting the entire system. The importance of modular design in secure systems is highlighted in \cite{ziller2021pysyft}. 3) Establish a platform for secure and anonymous reporting of vulnerabilities by users and developers. Incentivize responsible disclosure through recognition or rewards, encouraging active participation in the framework's security enhancement. This extends practices from open-source security communities as discussed in \cite{eghbal2020working}.

\subsection{Policy for Protecting Against Compromised Servers}
1) Transition from a centralized aggregator to a decentralized aggregation protocol utilizing secure multi-party computation (SMPC). This approach eliminates single points of failure and enhances security, as described in \cite{bonawitz2017practical}, who explore practical SMPC for privacy-preserving machine learning. 2) Implement robust security measures on servers, including intrusion detection systems, regular security updates, and adherence to best practices for server hardening. Regular penetration testing helps identify and address vulnerabilities. \cite{scarfone2009guide} provides guidelines on server hardening practices. 3) Utilize homomorphic encryption to allow the server to perform aggregation on encrypted model updates without accessing the underlying data. This ensures that even if the server is compromised, sensitive information remains protected, as explored in \cite{acar2018survey}, which surveys homomorphic encryption schemes. 4) Develop and maintain a detailed incident response plan that outlines procedures for detecting, reporting, and responding to server compromises. This includes immediate isolation of affected systems and communication protocols to inform clients. Guidelines for incident response are detailed in \cite{cichonski2012computer}.

\subsection{Policy for Defending Against External Adversaries}
1) Mandate the use of strong encryption protocols, such as TLS 1.3 with AES-256 encryption, for all communications between clients and servers. This protects data in transit from interception and tampering, following guidelines from \cite{mckay2017guidelines}. 2)  Implement protocols that provide forward secrecy and resistance to replay attacks, such as the Noise Protocol Framework. This enhances the security of communications against sophisticated external adversaries, as recommended in \cite{perrin2018noise}. 3) Deploy intrusion detection systems and integrate threat intelligence feeds to monitor for and respond to emerging threats. Regular updates based on current threat landscapes help maintain robust defenses. Strategies for threat intelligence are discussed in \cite{barnum2012standardizing}. 4) Conduct regular training sessions for personnel involved in the operation of the DPFedBank framework to ensure awareness of security best practices and protocols for handling potential security incidents. \cite{linckeinformation} highlights the importance of ongoing security awareness training. 

\subsection{Policy for Data Privacy Compliance}
Ensure that the DPFedBank framework complies with relevant data protection laws such as GDPR, CCPA, and other applicable regulations. This includes conducting Data Protection Impact Assessments (DPIAs), as recommended by the \cite{grutter2019data}. 2) Implement mechanisms to monitor and control the cumulative privacy loss (\(\epsilon\)) over time for each client. This prevents excessive data exposure and aligns with privacy accounting methods discussed in \cite{dwork2014algorithmic}, who elaborate on differential privacy foundations. 3) Clearly communicate the privacy guarantees and data handling practices to all participating institutions, ensuring transparency and building trust among collaborators. Transparency is emphasized in \cite{kleinberg2016inherent}, discussing fairness and accountability in AI systems.

\subsection{Policy for Inter-Client Trust and Collaboration}
1) Establish comprehensive legal agreements that define the rights, responsibilities, and liabilities of each participating institution. These agreements should cover data usage policies, confidentiality obligations, and dispute resolution mechanisms. Similar frameworks are used in blockchain consortia, as noted by \cite{brown2018corda}. 2) Create a governance structure comprising representatives from participating institutions to oversee the operation of the DPFedBank framework. This body would be responsible for making decisions on policy changes, security practices, and handling incidents. Governance models in collaborative networks are discussed in \cite{beck2017blockchain}. 3) Develop and adopt standardized protocols for data formats, encryption methods, and communication interfaces to ensure seamless interoperability among different institutions' systems. This approach is recommended by \cite{documentation2005information} for information security management systems.

\subsection{Policy for Deterring Data Inference Attacks}

1) Mandate the adjustment of noise levels in the LDP mechanism within the DPFedBank framework to address privacy risks associated with non-IID data, where unique patterns at individual institutions pose a threat. Institutions must fine-tune noise levels based on the data distribution characteristics at each client, thereby reducing the risk of exposing distinctive patterns. For instance, financial institutions with distinct customer behaviors must apply higher noise levels when perturbing local model updates to ensure these unique patterns are less detectable during aggregation, strengthening privacy guarantees across all clients.

2) Require the application of gradient compression in DPFedBank, where clients (e.g., banks or credit unions) locally train their models and transmit perturbed updates. Institutions must implement gradient compression techniques, such as quantization or sparsification \cite{haddadpour2021federated}, to reduce the amount of data transferred and lower the risk of exposing sensitive patterns in local model updates. This practice is essential for handling non-IID data, where overrepresented patterns could lead to data leakage. For example, a credit union should compress gradients before applying Local Differential Privacy (LDP) to minimize information leakage during transmission and enhance privacy-preserving mechanisms across clients with varying data distributions.

3) Implement Secure Multiparty Computation (SMC) as a standard to enable collaborative model training across financial institutions without revealing individual datasets. In the DPFedBank framework, SMC must be used to securely aggregate locally perturbed model updates, providing stronger privacy guarantees than relying solely on LDP. Institutions should adopt SMC-based protocols such as Chain-PPFL \cite{li2020privacy} to facilitate the exchange of masked gradients and mitigate data leakage risks. Advances in SMC protocols \cite{wang2017global} should be leveraged to ensure scalable secure computation, even in cases of non-IID data distributions.

\subsection{Policy for Mitigating Model Poisoning Attacks}

1) Enforce the integration of blockchain technology into the DPFedBank framework as a defense against model poisoning attacks. Institutions must leverage blockchain’s immutable nature to ensure that updates to the global model are transparent, verified, and securely stored, making tampering attempts easily traceable and detectable \cite{kalapaaking2023blockchain}. This policy is crucial for safeguarding models used in anti-money laundering (AML), credit risk assessments, and fraud detection. If anomalous or potentially poisoned updates are detected, blockchain-based records should facilitate immediate investigation and rejection of compromised data, ensuring that only valid updates are incorporated into the global model.

2) Adopt the Biscotti system for decentralized federated learning within DPFedBank. Institutions should utilize the Biscotti system \cite{shayan2020biscotti} to coordinate model updates using a blockchain-based approach, eliminating the need for a central aggregator. The Biscotti system must incorporate Proof-of-Federation (PoF), Multi-Krum (for filtering out anomalous updates), differentially private noise, and Shamir secret sharing to secure model updates and log the aggregation process on the blockchain. This decentralized approach is especially suited for financial institutions that prefer not to rely on centralized infrastructure due to privacy and regulatory concerns.

3) Implement Byzantine-resilient stochastic gradient descent (SGD) methods to defend against adversarial manipulation in federated learning. Financial institutions must integrate Byzantine-resilient SGD methods, such as HoldOut SGD \cite{azulay2020holdout} and Generalized Byzantine-tolerant SGD \cite{xie2018generalized}, to filter out outliers and poisoned gradients, ensuring the integrity of the global model. For instance, trusted branches within DPFedBank should evaluate gradient updates to confirm that only legitimate updates are accepted. These methods are particularly important for high-stakes applications like high-frequency trading and fraud detection, where data integrity is critical.

\subsection{Policy for Protecting Against Insider Based Attacks}
1) Establish role-based access controls (RBAC) to restrict potential insiders' access to financial information in models used for federated learning. The risk of insider misuse should be greatly decreased by limiting access to particular datasets and model parameters to those who have received explicit approval \cite{1416861}.
2) Implement continuous auditing in the federated learning environment.To conduct ongoing audits to monitor all model interactions and data access. By monitoring potential patterns, audit trails can detect potential insider threats. \cite{anomarticle}.
3) Automate the differential privacy (DP) noise injection process in order to potentially prevent insider manipulation of the differential privacy (DP) noise injection process. By automating this process it guarantees that any insider cannot change the noise settings to potentially weaken this privacy protection.\cite{10.1561/0400000042}
4) Conduct insider risk assessments to evaluate potential vulnerabilities that can be exploited by insiders. To limit the risk of insiders, periodic assessments can incorporate analysis of the system’s performance, employee behavior, and access patterns. \cite{INAYAT2024103068}

\subsection{Policy for Protecting against Meet in the Middle Attacks}
1)Implement a Data Encryption Strategy, this can be achieved by implementing by using a combination of homomorphic and end-to-end encryption to protect private financial information during federated learning. Homomorphic encryption permits calculations on encrypted data without disclosing its contents, and end-to-end encryption guarantees that data is protected during transit between organizations. This is a dual-layer approach where private financial data is not exposed during interception or during the collaborative model training\cite{app12020734}
2)Implement secure multiparty computation (SMPC) to prevent meet-in-the-middle attacks by allowing several organizations to collaborate on compute functions without disclosing their personal inputs. \cite{bonawitz2017practical}. 
3)Randomize model updates to introduce variance into the updates shared during model training. This unpredictability reduces the possibility of meet-in-the-middle attacks by preventing attackers from comparing communications between parties.
4)Implement integrity verification protocols, such as cryptographic hash algorithm. This is to guarantee the integrity of all data including model updates sent across organizations. Hash-based integrity verification techniques offer a barrier against possible meet-in-the-middle attackers and detect if there is any tampering during data transfer.\cite{Balasubramanian2024AFL}

\section{Conclusion}

The proposed policies and regulations are designed to significantly enhance the security, privacy, and trustworthiness of the DPFedBank framework, thereby positioning it as a leading solution for privacy-preserving federated learning within the financial sector. By incorporating stringent authentication measures, advanced encryption techniques, proactive vulnerability management, and collaborative governance structures, these policies effectively mitigate a wide spectrum of potential threats. This includes threats from malicious clients seeking unauthorized access, inherited system flaws that could expose vulnerabilities, compromised servers that could be exploited for data breaches, and sophisticated external adversaries attempting to undermine the system's integrity.

Furthermore, these measures work collectively to address both internal and external risks, ensuring that the framework is resilient against a variety of sophisticated attack vectors. The use of advanced encryption techniques guarantees that data remains secure during transit and storage, while proactive vulnerability management allows the system to anticipate and rectify security gaps before they can be exploited. Collaborative governance structures ensure that all stakeholders are aligned in their efforts to maintain high security standards, thereby reducing the risk of security lapses due to human error or miscommunication. Through this comprehensive approach, the DPFedBank framework is able to create a secure ecosystem that not only protects sensitive financial data but also supports regulatory compliance and fosters trust among all participants.

The integration of these policies not only fortifies the framework against current and emerging risks but also establishes a robust foundation for trust and cooperation among participating financial institutions. This foundation is instrumental in creating a cohesive network where stakeholders can share data and insights with confidence, knowing that their security and privacy needs are prioritized. Moreover, these enhancements foster a secure and transparent collaborative environment, encouraging financial entities to leverage federated learning without compromising sensitive data. By ensuring that all participating institutions adhere to a unified set of high standards, these policies facilitate interoperability and consistency in data handling practices across the board. Such trust is pivotal for ensuring compliance with regulatory requirements, streamlining audit processes, and maintaining public confidence in data-driven innovation in banking and finance. Furthermore, the commitment to transparency helps address the concerns of both customers and regulatory bodies, making it easier for institutions to demonstrate adherence to best practices and legal obligations.

Ultimately, the strengthened DPFedBank framework enables secure, efficient, and privacy-preserving data sharing across institutions, driving innovation while upholding the highest standards of data protection. This positions financial institutions to fully leverage the capabilities of collaborative machine learning, resulting in more informed decision-making, enhanced operational efficiency, and superior customer services across the industry.

\section*{Acknowledgment}
\thanks{$^{\dagger}$ Peilin He, Chenkai Lin, and Isabella Montoya contributed equally to this work, last name in alphabetical order.}



%
%
%
%
\bibliographystyle{IEEEtran}
\bibliography{ref}

\begin{thebibliography}{10}
\providecommand{\url}[1]{#1}
\csname url@samestyle\endcsname
\providecommand{\newblock}{\relax}
\providecommand{\bibinfo}[2]{#2}
\providecommand{\BIBentrySTDinterwordspacing}{\spaceskip=0pt\relax}
\providecommand{\BIBentryALTinterwordstretchfactor}{4}
\providecommand{\BIBentryALTinterwordspacing}{\spaceskip=\fontdimen2\font plus
\BIBentryALTinterwordstretchfactor\fontdimen3\font minus \fontdimen4\font\relax}
\providecommand{\BIBforeignlanguage}[2]{{%
\expandafter\ifx\csname l@#1\endcsname\relax
\typeout{** WARNING: IEEEtran.bst: No hyphenation pattern has been}%
\typeout{** loaded for the language `#1'. Using the pattern for}%
\typeout{** the default language instead.}%
\else
\language=\csname l@#1\endcsname
\fi
#2}}
\providecommand{\BIBdecl}{\relax}
\BIBdecl

\bibitem{mammen2021federated}
P.~M. Mammen, ``Federated learning: Opportunities and challenges,'' \emph{arXiv preprint arXiv:2101.05428}, 2021.

\bibitem{li2020review}
L.~Li, Y.~Fan, M.~Tse, and K.-Y. Lin, ``A review of applications in federated learning,'' \emph{Computers \& Industrial Engineering}, vol. 149, p. 106854, 2020.

\bibitem{wei2020federated}
K.~Wei, J.~Li, M.~Ding, C.~Ma, H.~H. Yang, F.~Farokhi, S.~Jin, T.~Q. Quek, and H.~V. Poor, ``Federated learning with differential privacy: Algorithms and performance analysis,'' \emph{IEEE transactions on information forensics and security}, vol.~15, pp. 3454--3469, 2020.

\bibitem{bouacida2021vulnerabilities}
N.~Bouacida and P.~Mohapatra, ``Vulnerabilities in federated learning,'' \emph{IEEE Access}, vol.~9, pp. 63\,229--63\,249, 2021.

\bibitem{kairouz2021advances}
P.~Kairouz, H.~B. McMahan, B.~Avent, A.~Bellet, M.~Bennis, A.~N. Bhagoji, K.~Bonawitz, Z.~Charles, G.~Cormode, R.~Cummings \emph{et~al.}, ``Advances and open problems in federated learning,'' \emph{Foundations and trends{\textregistered} in machine learning}, vol.~14, no. 1--2, pp. 1--210, 2021.

\bibitem{FederatedLearning}
J.~Konečný, B.~McMahan, and D.~Ramage, ``Federated optimization:distributed optimization beyond the datacenter,'' 11 2015.

\bibitem{Han2023FederatedLD}
\BIBentryALTinterwordspacing
L.~Han, D.~Fan, J.~Liu, and W.~Du, ``Federated learning differential privacy preservation method based on differentiated noise addition,'' \emph{2023 8th International Conference on Cloud Computing and Big Data Analytics (ICCCBDA)}, pp. 285--289, 2023. [Online]. Available: \url{https://api.semanticscholar.org/CorpusID:259279306}
\BIBentrySTDinterwordspacing

\bibitem{Shen2023ADP}
\BIBentryALTinterwordspacing
Z.~Shen, S.~He, H.~Wang, P.~Liu, K.~Liu, and F.~Lian, ``A differential privacy budget allocation method combining privacy security level,'' \emph{J. Commun. Inf. Networks}, vol.~8, pp. 90--98, 2023. [Online]. Available: \url{https://api.semanticscholar.org/CorpusID:257849270}
\BIBentrySTDinterwordspacing

\bibitem{Drainakis2020FederatedVC}
\BIBentryALTinterwordspacing
G.~Drainakis, K.~V. Katsaros, P.~Pantazopoulos, V.~Sourlas, and A.~J. Amditis, ``Federated vs. centralized machine learning under privacy-elastic users: A comparative analysis,'' \emph{2020 IEEE 19th International Symposium on Network Computing and Applications (NCA)}, pp. 1--8, 2020. [Online]. Available: \url{https://api.semanticscholar.org/CorpusID:230997062}
\BIBentrySTDinterwordspacing

\bibitem{Nilsson2018APE}
\BIBentryALTinterwordspacing
A.~Nilsson, S.~Smith, G.~Ulm, E.~Gustavsson, and M.~Jirstrand, ``A performance evaluation of federated learning algorithms,'' \emph{Proceedings of the Second Workshop on Distributed Infrastructures for Deep Learning}, 2018. [Online]. Available: \url{https://api.semanticscholar.org/CorpusID:53711711}
\BIBentrySTDinterwordspacing

\bibitem{DBLP:journals/corr/abs-1911-11815}
\BIBentryALTinterwordspacing
M.~Fang, X.~Cao, J.~Jia, and N.~Z. Gong, ``Local model poisoning attacks to byzantine-robust federated learning,'' \emph{CoRR}, vol. abs/1911.11815, 2019. [Online]. Available: \url{http://arxiv.org/abs/1911.11815}
\BIBentrySTDinterwordspacing

\bibitem{PrivacyHarm}
M.~Calo, ``The boundaries of privacy harm,'' \emph{M. Ryan Calo}, vol.~86, 07 2010.

\bibitem{GDPR}
P.~Voigt and A.~Bussche, \emph{The EU General Data Protection Regulation (GDPR): A Practical Guide}, 01 2017.

\bibitem{PrivacyLaw}
D.~Solove, ``A brief history of information privacy law,'' 07 2006.

\bibitem{10.1145/3338501.3357370}
\BIBentryALTinterwordspacing
S.~Truex, N.~Baracaldo, A.~Anwar, T.~Steinke, H.~Ludwig, R.~Zhang, and Y.~Zhou, ``A hybrid approach to privacy-preserving federated learning,'' in \emph{Proceedings of the 12th ACM Workshop on Artificial Intelligence and Security}, ser. AISec'19.\hskip 1em plus 0.5em minus 0.4em\relax New York, NY, USA: Association for Computing Machinery, 2019, p. 1–11. [Online]. Available: \url{https://doi.org/10.1145/3338501.3357370}
\BIBentrySTDinterwordspacing

\bibitem{shokri2015privacy}
R.~Shokri and V.~Shmatikov, ``Privacy-preserving deep learning,'' in \emph{Proceedings of the 22nd ACM SIGSAC conference on computer and communications security}, 2015, pp. 1310--1321.

\bibitem{fredrikson2015model}
M.~Fredrikson, S.~Jha, and T.~Ristenpart, ``Model inversion attacks that exploit confidence information and basic countermeasures,'' in \emph{Proceedings of the 22nd ACM SIGSAC conference on computer and communications security}, 2015, pp. 1322--1333.

\bibitem{dwork2014algorithmic}
C.~Dwork, A.~Roth \emph{et~al.}, ``The algorithmic foundations of differential privacy,'' \emph{Foundations and Trends{\textregistered} in Theoretical Computer Science}, vol.~9, no. 3--4, pp. 211--407, 2014.

\bibitem{abadi2016deep}
M.~Abadi, A.~Chu, I.~Goodfellow, H.~B. McMahan, I.~Mironov, K.~Talwar, and L.~Zhang, ``Deep learning with differential privacy,'' in \emph{Proceedings of the 2016 ACM SIGSAC conference on computer and communications security}, 2016, pp. 308--318.

\bibitem{bagdasaryan2020backdoor}
E.~Bagdasaryan, A.~Veit, Y.~Hua, D.~Estrin, and V.~Shmatikov, ``How to backdoor federated learning,'' in \emph{International conference on artificial intelligence and statistics}.\hskip 1em plus 0.5em minus 0.4em\relax PMLR, 2020, pp. 2938--2948.

\bibitem{geyer2017differentially}
R.~C. Geyer, T.~Klein, and M.~Nabi, ``Differentially private federated learning: A client level perspective,'' \emph{arXiv preprint arXiv:1712.07557}, 2017.

\bibitem{lyu2022privacy}
L.~Lyu, H.~Yu, X.~Ma, C.~Chen, L.~Sun, J.~Zhao, Q.~Yang, and S.~Y. Philip, ``Privacy and robustness in federated learning: Attacks and defenses,'' \emph{IEEE transactions on neural networks and learning systems}, 2022.

\bibitem{sandhu1993lattice}
R.~S. Sandhu, ``Lattice-based access control models,'' \emph{Computer}, vol.~26, no.~11, pp. 9--19, 1993.

\bibitem{samarati2000access}
P.~Samarati and S.~C. De~Vimercati, ``Access control: Policies, models, and mechanisms,'' in \emph{International school on foundations of security analysis and design}.\hskip 1em plus 0.5em minus 0.4em\relax Springer, 2000, pp. 137--196.

\bibitem{mcmahan2017communication}
B.~McMahan, E.~Moore, D.~Ramage, S.~Hampson, and B.~A. y~Arcas, ``Communication-efficient learning of deep networks from decentralized data,'' in \emph{Artificial intelligence and statistics}.\hskip 1em plus 0.5em minus 0.4em\relax PMLR, 2017, pp. 1273--1282.

\bibitem{duchi2013local}
J.~C. Duchi, M.~I. Jordan, and M.~J. Wainwright, ``Local privacy and statistical minimax rates,'' in \emph{2013 IEEE 54th annual symposium on foundations of computer science}.\hskip 1em plus 0.5em minus 0.4em\relax IEEE, 2013, pp. 429--438.

\bibitem{triastcyn2019federated}
A.~Triastcyn and B.~Faltings, ``Federated learning with bayesian differential privacy,'' in \emph{2019 IEEE International Conference on Big Data (Big Data)}.\hskip 1em plus 0.5em minus 0.4em\relax IEEE, 2019, pp. 2587--2596.

\bibitem{dwork2006differential}
C.~Dwork, ``Differential privacy,'' in \emph{International colloquium on automata, languages, and programming}.\hskip 1em plus 0.5em minus 0.4em\relax Springer, 2006, pp. 1--12.

\bibitem{zhao2018federated}
Y.~Zhao, M.~Li, L.~Lai, N.~Suda, D.~Civin, and V.~Chandra, ``Federated learning with non-iid data,'' \emph{arXiv preprint arXiv:1806.00582}, 2018.

\bibitem{zhu2021federated}
H.~Zhu, J.~Xu, S.~Liu, and Y.~Jin, ``Federated learning on non-iid data: A survey,'' \emph{Neurocomputing}, vol. 465, pp. 371--390, 2021.

\bibitem{ma2022state}
X.~Ma, J.~Zhu, Z.~Lin, S.~Chen, and Y.~Qin, ``A state-of-the-art survey on solving non-iid data in federated learning,'' \emph{Future Generation Computer Systems}, vol. 135, pp. 244--258, 2022.

\bibitem{singh2023fair}
A.~K. Singh, A.~Blanco-Justicia, and J.~Domingo-Ferrer, ``Fair detection of poisoning attacks in federated learning on non-iid data,'' \emph{Data Mining and Knowledge Discovery}, vol.~37, no.~5, pp. 1998--2023, 2023.

\bibitem{shen2023privacy}
X.~Shen, Y.~Liu, F.~Li, and C.~Li, ``Privacy-preserving federated learning against label-flipping attacks on non-iid data,'' \emph{IEEE Internet of Things Journal}, vol.~11, no.~1, pp. 1241--1255, 2023.

\bibitem{dwork2009differential}
C.~Dwork and J.~Lei, ``Differential privacy and robust statistics,'' in \emph{Proceedings of the forty-first annual ACM symposium on Theory of computing}, 2009, pp. 371--380.

\bibitem{li2020federated}
T.~Li, A.~K. Sahu, M.~Zaheer, M.~Sanjabi, A.~Talwalkar, and V.~Smith, ``Federated optimization in heterogeneous networks,'' \emph{Proceedings of Machine learning and systems}, vol.~2, pp. 429--450, 2020.

\bibitem{bonawitz2017practical}
K.~Bonawitz, V.~Ivanov, B.~Kreuter, A.~Marcedone, H.~B. McMahan, S.~Patel, D.~Ramage, A.~Segal, and K.~Seth, ``Practical secure aggregation for privacy-preserving machine learning,'' in \emph{proceedings of the 2017 ACM SIGSAC Conference on Computer and Communications Security}, 2017, pp. 1175--1191.

\bibitem{sabt2015trusted}
M.~Sabt, M.~Achemlal, and A.~Bouabdallah, ``Trusted execution environment: What it is, and what it is not,'' in \emph{2015 IEEE Trustcom/BigDataSE/Ispa}, vol.~1.\hskip 1em plus 0.5em minus 0.4em\relax IEEE, 2015, pp. 57--64.

\bibitem{costan2016intel}
V.~Costan, ``Intel sgx explained,'' \emph{IACR Cryptol, EPrint Arch}, 2016.

\bibitem{hunt2018chiron}
T.~Hunt, C.~Song, R.~Shokri, V.~Shmatikov, and E.~Witchel, ``Chiron: Privacy-preserving machine learning as a service,'' \emph{arXiv preprint arXiv:1803.05961}, 2018.

\bibitem{ziller2021pysyft}
A.~Ziller, A.~Trask, A.~Lopardo, B.~Szymkow, B.~Wagner, E.~Bluemke, J.-M. Nounahon, J.~Passerat-Palmbach, K.~Prakash, N.~Rose \emph{et~al.}, ``Pysyft: A library for easy federated learning,'' \emph{Federated Learning Systems: Towards Next-Generation AI}, pp. 111--139, 2021.

\bibitem{eghbal2020working}
N.~Eghbal, \emph{Working in public: the making and maintenance of open source software}.\hskip 1em plus 0.5em minus 0.4em\relax Stripe Press, 2020.

\bibitem{scarfone2009guide}
K.~Scarfone, \emph{Guide to general server security: Recommendations of the national institute of standards and technology}.\hskip 1em plus 0.5em minus 0.4em\relax Diane Publishing, 2009.

\bibitem{acar2018survey}
A.~Acar, H.~Aksu, A.~S. Uluagac, and M.~Conti, ``A survey on homomorphic encryption schemes: Theory and implementation,'' \emph{ACM Computing Surveys (Csur)}, vol.~51, no.~4, pp. 1--35, 2018.

\bibitem{cichonski2012computer}
P.~Cichonski, T.~Millar, T.~Grance, K.~Scarfone \emph{et~al.}, ``Computer security incident handling guide,'' \emph{NIST Special Publication}, vol. 800, no.~61, pp. 1--147, 2012.

\bibitem{mckay2017guidelines}
K.~McKay and D.~Cooper, ``Guidelines for the selection, configuration, and use of transport layer security (tls) implementations,'' National Institute of Standards and Technology, Tech. Rep., 2017.

\bibitem{perrin2018noise}
T.~Perrin, ``The noise protocol framework,'' \emph{noiseprotocol, Protocol Revision}, vol.~34, 2018.

\bibitem{barnum2012standardizing}
S.~Barnum, ``Standardizing cyber threat intelligence information with the structured threat information expression (stix),'' \emph{Mitre Corporation}, vol.~11, pp. 1--22, 2012.

\bibitem{linckeinformation}
S.~Lincke, ``Information security planning.''

\bibitem{grutter2019data}
B.~J. Gr{\"u}tter and B.~Schneider, ``Data protection impact assessment guidelines in the context of the general data protection regulation,'' 2019.

\bibitem{kleinberg2016inherent}
J.~Kleinberg, S.~Mullainathan, and M.~Raghavan, ``Inherent trade-offs in the fair determination of risk scores,'' \emph{arXiv preprint arXiv:1609.05807}, 2016.

\bibitem{brown2018corda}
R.~G. Brown, ``The corda platform: An introduction,'' \emph{Retrieved}, vol.~27, p. 2018, 2018.

\bibitem{beck2017blockchain}
R.~Beck, M.~Avital, M.~Rossi, and J.~B. Thatcher, ``Blockchain technology in business and information systems research,'' pp. 381--384, 2017.

\bibitem{documentation2005information}
T.~Documentation and C.~LOGICAL, ``Information technology--security techniques--information security management systems--requirements,'' 2005.

\bibitem{haddadpour2021federated}
F.~Haddadpour, M.~M. Kamani, A.~Mokhtari, and M.~Mahdavi, ``Federated learning with compression: Unified analysis and sharp guarantees,'' in \emph{International Conference on Artificial Intelligence and Statistics}.\hskip 1em plus 0.5em minus 0.4em\relax PMLR, 2021, pp. 2350--2358.

\bibitem{li2020privacy}
Y.~Li, Y.~Zhou, A.~Jolfaei, D.~Yu, G.~Xu, and X.~Zheng, ``Privacy-preserving federated learning framework based on chained secure multiparty computing,'' \emph{IEEE Internet of Things Journal}, vol.~8, no.~8, pp. 6178--6186, 2020.

\bibitem{wang2017global}
X.~Wang, S.~Ranellucci, and J.~Katz, ``Global-scale secure multiparty computation,'' in \emph{Proceedings of the 2017 ACM SIGSAC Conference on Computer and Communications Security}, 2017, pp. 39--56.

\bibitem{kalapaaking2023blockchain}
A.~P. Kalapaaking, I.~Khalil, and X.~Yi, ``Blockchain-based federated learning with smpc model verification against poisoning attack for healthcare systems,'' \emph{IEEE Transactions on Emerging Topics in Computing}, vol.~12, no.~1, pp. 269--280, 2023.

\bibitem{shayan2020biscotti}
M.~Shayan, C.~Fung, C.~J. Yoon, and I.~Beschastnikh, ``Biscotti: A blockchain system for private and secure federated learning,'' \emph{IEEE Transactions on Parallel and Distributed Systems}, vol.~32, no.~7, pp. 1513--1525, 2020.

\bibitem{azulay2020holdout}
S.~Azulay, L.~Raz, A.~Globerson, T.~Koren, and Y.~Afek, ``Holdout sgd: Byzantine tolerant federated learning,'' \emph{arXiv preprint arXiv:2008.04612}, 2020.

\bibitem{xie2018generalized}
C.~Xie, O.~Koyejo, and I.~Gupta, ``Generalized byzantine-tolerant sgd,'' \emph{arXiv preprint arXiv:1802.10116}, 2018.

\bibitem{1416861}
E.~Bertino and R.~Sandhu, ``Database security - concepts, approaches, and challenges,'' \emph{IEEE Transactions on Dependable and Secure Computing}, vol.~2, no.~1, pp. 2--19, 2005.

\bibitem{anomarticle}
C.~Zhang, S.~Yang, L.-F. Mao, and H.~Ning, ``Anomaly detection and defense techniques in federated learning: a comprehensive review,'' \emph{Artificial Intelligence Review}, vol.~57, 05 2024.

\bibitem{10.1561/0400000042}
\BIBentryALTinterwordspacing
C.~Dwork and A.~Roth, ``The algorithmic foundations of differential privacy,'' \emph{Found. Trends Theor. Comput. Sci.}, vol.~9, no. 3–4, p. 211–407, Aug. 2014. [Online]. Available: \url{https://doi.org/10.1561/0400000042}
\BIBentrySTDinterwordspacing

\bibitem{INAYAT2024103068}
\BIBentryALTinterwordspacing
U.~Inayat, M.~Farzan, S.~Mahmood, M.~F. Zia, S.~Hussain, and F.~Pallonetto, ``Insider threat mitigation: Systematic literature review,'' \emph{Ain Shams Engineering Journal}, p. 103068, 2024. [Online]. Available: \url{https://www.sciencedirect.com/science/article/pii/S209044792400443X}
\BIBentrySTDinterwordspacing

\bibitem{app12020734}
\BIBentryALTinterwordspacing
J.~Park and H.~Lim, ``Privacy-preserving federated learning using homomorphic encryption,'' \emph{Applied Sciences}, vol.~12, no.~2, 2022. [Online]. Available: \url{https://www.mdpi.com/2076-3417/12/2/734}
\BIBentrySTDinterwordspacing

\bibitem{Balasubramanian2024AFL}
\BIBentryALTinterwordspacing
V.~Balasubramanian, M.~Aloqaily, and M.~Guizani, ``A federated learning secure encryption framework for autonomous systems,'' \emph{ICC 2024 - IEEE International Conference on Communications}, pp. 2197--2203, 2024. [Online]. Available: \url{https://api.semanticscholar.org/CorpusID:271939804}
\BIBentrySTDinterwordspacing

\end{thebibliography}

\end{document}